\documentclass[useAMS,usenatbib]{mn2e}
\usepackage{graphicx}
\usepackage{amssymb}
\usepackage{rotating}
\usepackage[export]{adjustbox}

\newcommand{\MC}{\multicolumn}
\newcommand{\kms}{km~s$^{-1}$}

\newcommand{\HI}{H{\sc i}}
\newcommand{\HII}{H{\sc ii}}
\newcommand{\sunn}{$_{\odot}$}

\title[UGC~3672: A gas-rich merging void triplet]
{UGC~3672: An unusual merging triplet of gas-rich galaxies in the Lynx-Cancer void}
\author[J.N.~Chengalur, S.A.~Pustilnik, E.S.~Egorova ]
{J.N.~Chengalur,$^{1}$\thanks{E-mail: chengalur@ncra.tifr.res.in
(JNC)} S.A.~Pustilnik,$^{2}$\thanks{E-mail: sap@sao.ru (SAP)}
E.S.~Egorova$^{3}$\\
$^1$ National Center for Radio Astrophysics, TIFR, Pune, India \\
$^2$ Special Astrophysical Observatory of RAS, Nizhnij Arkhyz,
  Karachai-Circassia 369167, Russia\\
$^3$ Sternberg Astronomical Institute, Lomonosov Moscow State University,
  13 Universitetsky pr., Moscow, 119991, Russia
}

\begin{document}

\label{firstpage}

\date{Accepted October 11, 2016? Received March 3, 2016}

%\pagerange{\pageref{firstpage}--\pageref{lastpage}} \pubyear{2016}

\maketitle

\begin{abstract} 

We present \HI\ 21cm and optical observations of UGC~3672 which is located near 
the centre of the nearby Lynx-Cancer void. We find that UGC~3672 consists of 
an approximately linearly aligned triplet of gas rich dwarfs with large scale 
velocity continuity along the triplet axis. The faintest component of the 
triplet is extremely gas-rich M$_{\rm HI}/L_{\rm B} \sim 17$) and also extremely 
metal deficient ($12+\log(O/H) \sim 7.0$). The metallicity of this dwarf is 
close to the 'floor' observed in star forming galaxies. Low resolution \HI\
images show that the galaxy triplet is located inside a common \HI\ envelope, 
with  fairly regular, disk like kinematics. At high angular resolution however,
the gas is found to be confined to several filamentary tidal tails and  
bridges. The linear alignment of the galaxies, along with the velocity 
continuity that we observe, is consistent with the galaxies lying along a 
filament. We argue that the location of this highly unusual system in an 
extremely low density environment is not a coincidence, but is a consequence of
structure formation proceeding more slowly and also probing smaller scales 
than in regions with average  density. Our observations also indicate that
wet mergers of galaxies flowing along filaments is a possible pathway for
the formation of gas rich disks. The UGC~3672 system provides an
interesting opportunity to study the kind of interactions typical between
high redshift  extremely gas rich unevolved small systems that lie at base
of the hierarchical galaxy  formation model. 

\end{abstract}

\begin{keywords}
% galaxies: dwarf -- (7 keywords is too much! need to choose)
galaxies: evolution -- galaxies: dwarf --
galaxies: interactions -- galaxies: individual: UGC~3672, UGC~3672A --
radio lines: galaxies -- cosmology: large-scale structure of Universe
\end{keywords}

\section[]{Introduction}
\label{sec:intro}
\setcounter{figure}{0}

The galaxy distribution in the nearby universe has been known for several 
decades to contain large regions almost completely devoid of bright
galaxies \citep{joeveer78,kirchner81}. Interest in these so called ``voids''
has been increasing in the recent years for a variety of reasons. They 
are a prominent feature of the cosmic web. Indeed structure formation 
in the universe can be regarded as being driven by matter flowing out
of the voids and into walls, and then along filaments and finally into
the clusters and super-clusters found at the intersection of the filaments.
The shape and abundance of voids are sensitive to cosmological parameters 
\citep[see e.g.][]{park07,biswas10} and models of gravity
\citep[][]{Cai15,Zivick15}. As such comparison of the properties of voids
observed in redshift surveys with numerical simulations are an important
independent way of constraining the values of dark energy density,
\citep[e.g.]{sutter12} and possibly even fundamental physics. Further,
although voids were first identified as regions devoid of galaxies, numerical
simulations as well as observations indicate that voids are not completely
empty. The study of galaxies in voids gives one the opportunity to study the
effect of environment on the formation and evolution of galaxies.

Numerical simulations predict that the mass function of dark matter halos
is strongly influenced by the environment. In void regions its is predicted
that the mass function is steeper than in the walls and filaments, i.e.
the number of massive halos is depressed compared to that of the low
mass halos \citep[e.g.][]{gottlober03}. There is also a spatial segregation,
in that the more massive halos are expected to be found nearer the walls which
define the void, whereas the less massive halos could be found distributed
throughout the void \citep[e.g.]{peebles01}. Observations of void galaxies
identified using the SDSS survey do indeed provide observational support for
the prediction that the interior of voids are dominated by fainter bluer
galaxies \citep[see e.g.][]{rojas04,hoyle05,hoyle12, liu15} with higher
specific star formation rates \citep{rojas05,moorman16} and that there does
appear to be a tendency for dwarf galaxies to be dominant as one goes towards
the centre of voids \citep{hoyle12}. Similarly \HI\ observations of nearby
voids find evidence for the \HI\ mass function moving to a lower
characteristic mass in the interior of voids \citep{moorman14}. Some tension
between observation and theory however still remains. For example,
the total number of dwarfs found in the interior of voids is significantly
smaller than the number of low mass dark matter halos predicted by
simulations \citep{grogin99,hoeft06}.

Although voids are low density regions produced by the streaming of matter
out of voids and towards the higher density walls and filaments,
\cite[e.g.]{Dubinski93,Sheth04} they are not devoid of internal structure.
Voids contain sub-voids, which are themselves delineated by walls, filaments
and high density regions at the intersection of the filaments
\citep[e.g.][]{sahni94,aragon-calvo13}. Indeed to first order structure
formation in voids is similar to structure formation in a low density
universe \citep{goldberg04}.  Structure formation 
in voids proceeds slower than in high density regions, and because the 
interiors of voids are expanding faster than the average expansion rate, 
structures in voids are also expanded compared to the mean. Voids hence give 
one the possibility of looking at both the earlier stages 
of structure formation as well as probing smaller scales of the power 
spectrum than is probed by structures in dense regions, and as such,
voids can be regarded as ``cosmic microscopes and time machines'' 
\citep{aragon-calvo13}. Since the effective expansion rate increases towards 
the central low density regions of the voids these effects get more 
pronounced the closer one gets to the centre of the void.

For all of the above reasons it is interesting to study galaxies found inside
the voids. Recent surveys have found several unusual objects located inside
voids. For e.g. \citet{kreckel11} discuss the case of KK246, a dwarf
galaxy with an extended and disturbed disk lying in the Tully void, and 
\citet{beygu13} discuss the case of an approximately linear triplet of rich
galaxies found as part of their Void Galaxy Survey (VGS). They suggest that 
this linear arrangement may arise due to material flow along a filament 
inside the void. Numerical simulations of \citet{rieder13} provide some
support for this suggestion.

We have for some time been conducting a detailed survey of the galaxies
in a very nearby void, the Lynx-Cancer void \citep{pustilnik11}. The
Lynx-Cancer void is located at the edge of the Local Volume - its
centre only $\sim 18$ Mpc~distant. Since this is a nearby void, it
gives one the opportunity to study galaxies to a much fainter mass and 
luminosity limit than have been done by mentioned above authors
in more distant voids. The void galaxy
sample of \citet{pustilnik11} is estimated to be nearly complete to 
M$_{\rm B} < -14$~mag, and contains galaxies as faint as
M$_{\rm B} ~ -11.9$~mag. The updated version of the sample
(in preparation) contains galaxies down to M$_{\rm B} ~ -9.6$~mag.
The galaxies in the Lynx-Cancer void tend to be metal deficient compared
to the population of a denser environment
%\citep[][and Pustilnik,Perepelitsyna \& Kniazev, 2016, submitted]{pustilniketal11a}
\citep{pustilniketal11a,Paper7}
and the void contains some of the most metal-poor and gas-rich galaxies known
\citep{DDO68,J0926,pustilniketal11b,IT07,chengalur13}.

A survey of the \HI\ content of the void galaxies is presented in
% Pustilnik \& Martin (2016, submitted),
\citet{Paper6}, and \HI\ imaging of a selected
subset of gas-rich galaxies is also being undertaken with the GMRT.
The ongoing GMRT observations
have already led to the discovery of several interesting systems. These 
include a triplet of extremely gas-rich faint dwarf galaxies J0723+36,
once again with an approximately linear alignment \citep{chengalur13} and
the UGC~4722
system \citep{chengalur15}, which is  found to consist of an interacting 
pair of metal-poor galaxies joined by blue plume which appears to consist
almost entirely of young stars. One more void triplet is very metal-poor
galaxy DDO~68, in which the two more massive components already merged
\citep{ekta08}, while the third, much fainter component, DDO~68C at $\sim$42~kpc,
shows traces of pulled out \HI\ gas \citep{cannon14}.
In this paper we discuss UGC~3672, a
galaxy which lies in the central 8\%  of the void volume.
The galaxy density in the neighbourhood of UGC~3672 has been estimated to
be $\sim$10 times lower than the mean density.

The rest of this paper is arranged as follows. The GMRT observations and the
optical photometry are presented in Sec.~\ref{sec:obs}. The results of the
observations are presented in Sec.~\ref{sec:results} and discussed in
Sec.~\ref{sec:dis}.  Throughout the paper, we follow \citet{pustilnik11}
in assuming a distance of 16.9~Mpc to UGC~3672 (scale 82~pc/\arcsec).
The latter distance, in turn,
takes into account the large peculiar velocity of $\Delta V \sim -$270~\kms\
in this region (after \citet{tully08}).

\section[]{Observations and data reduction}
\label{sec:obs}

\subsection{GMRT data}
\label{ssec:gmrt}

GMRT \HI\ 21~cm observations of UGC~3672 were carried out on 8 and 9 September
2015, with a total on source time of $\sim$5.3 hours.  The correlator was
configured to a total bandwidth of 4.17~MHz ($\sim$890~\kms) divided into 
512 channels (or a velocity resolution of 1.74~\kms, the data have not been
hanning smoothed since the GMRT FX correlator's spectral leakage is small)
centred at the heliocentric redshift of the galaxy. The initial flagging and
calibration were carried out using the FLAGCAL pipeline \citep{flagcal,
flagcal1}, and the subsequent processing was done using the AIPS package. A
continuum image was made using the line free channels and used for
self-calibration. The self calibration solutions were then applied to the
line visibilities, and the continuum emission subtracted out using the task
UVSUB. Images were then made at a variety of resolutions (by applying
different UV tapers) using the task IMAGR, and residual continuum subtracted
out using the task IMLIN. Images of the integrated \HI\ emission and
the \HI\ velocity field were made using the task MOMNT.

\begin{table}
\caption{Parameters of the GMRT observations}
\label{tab:obspar}
\begin{tabular}{ll}
\hline
     & UGC~3672  \\
\hline
Date of observations     & 2015 Sep 8,9  \\
Field center R.A.(2000)  &07$^{h}$06$^{m}$27.5$^{s}$   \\
Field center Dec.(2000)  &+30$^{o}$19$^{'}$19.0$^{"}$    \\
Central Velocity (\kms)  & 994.0   \\
Time on-source  (h)      &$\sim$5.3  \\
Number of channels       & 512 \\
Channel separation (\kms)& $\sim$1.73 \\
Flux Calibrators         & 3C48,3C286 \\
Phase Calibrators        & 0741+312 \\
\hline
\hline
\end{tabular}
\end {table}

\subsection{Photometric data}
\label{ssec:photo}

Photometry of the UGC~3672 system was done based on the $u,g,r,i$ images
taken from the Sloan Digital Sky Survey DR12 \citep{dr12}, as well as
KPNO 0.9m B band and H-$\alpha$ images from \cite{vanzee00}. For the SDSS releases
later than DR8 all imaging data have been recalibrated using the improved 
procedure described in \citet{padmanabhan08}. DR12 provides reduced frames 
after sky-subtraction and this calibration; these can be directly used to
derive the photometric fluxes. SDSS fluxes are expressed in a linear scale
with units of nanomaggies, from which the Pogson magnitude can be estimated 
as $m = 22.5^{m} - 2.5 \log (f)$, where $f$ is the flux expressed in 
nanomaggies. We convert our integrated $g$- and $r$ band magnitudes into 
$B$ band magnitudes  $B_{\rm tot}$  using the transformation formula proposed
by Lupton et al. (2005): $B = g + 0.313(g - r) + 0.2271$; $\sigma$ = 0.0107.
In the case of KPNO 0.9m images, the sky subtracted and calibrated frames were
downloaded from NED. The H$\alpha$ and $B$ band magnitudes were computed
using the transform formulae provided in the FITS headers. 

\section[]{Results}
\label{sec:results}

\subsection[]{Low Resolution Data}
\label{ssec:lowres}

\begin{figure*}
  \begin{tabular}{p{0.5\textwidth} p{0.5\textwidth}}
    \vspace{0pt} \includegraphics[width=11.0cm,clip]{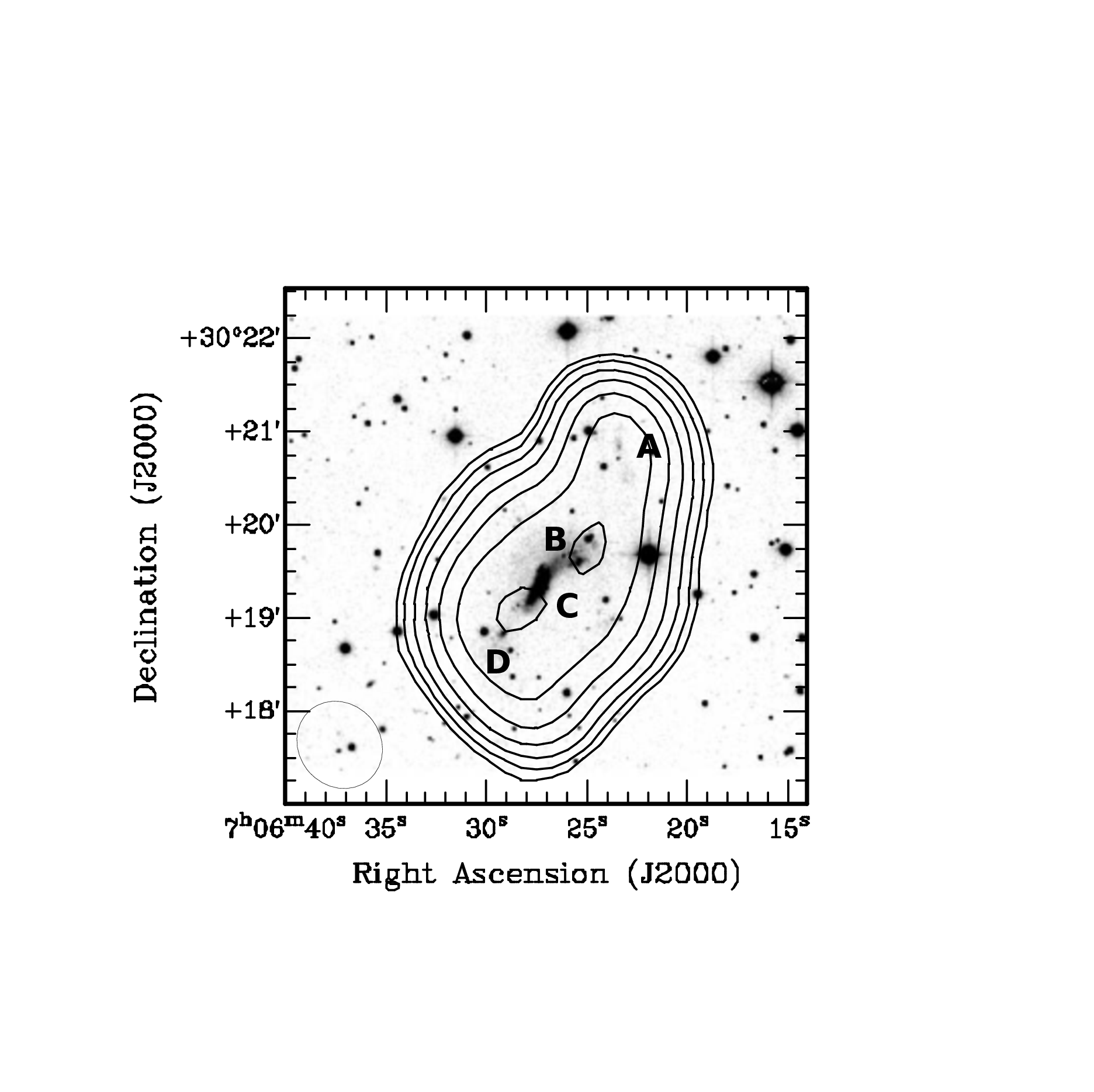} &
    \vspace{0pt} \includegraphics[width=11.0cm,clip]{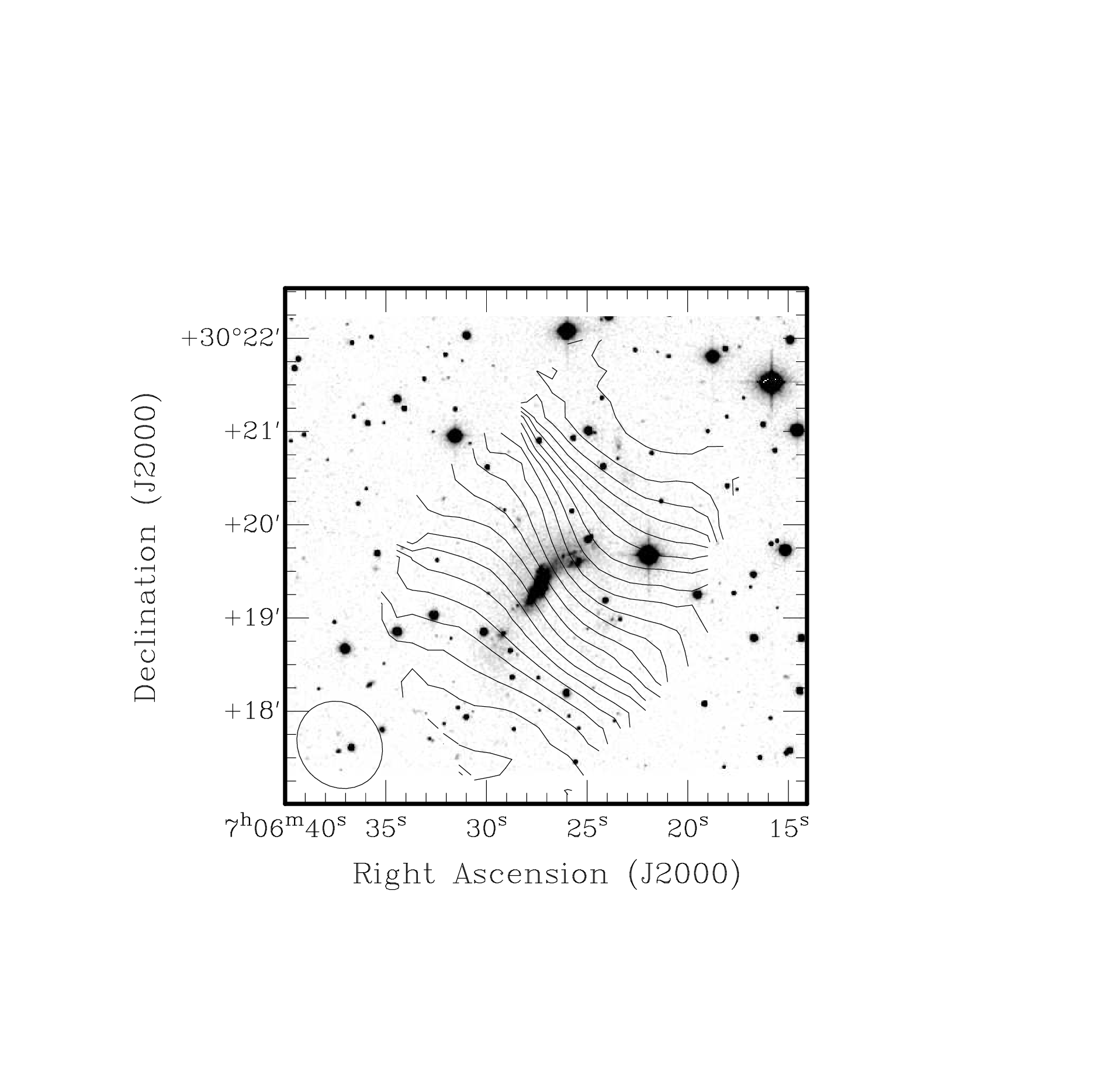}
  \end{tabular}
\caption{ {\bf Left Panel:} An overlay of the integrated \HI\ emission
(contours) at a resolution of 58\farcs5 $\times$ 52\farcs6 on a KPNO 0.9m
$B$ band optical
image (greyscales) of the UGC~3672 system. The \HI\ contours start at
a column density of $5.5 \times 10^{19}$~atoms/cm$^{-2}$ and are spaced a 
factor of 1.5 apart. See the text for the labelling of the different
components. The beam is shown in the bottom left.
{\bf Right Panel:} Velocity field obtained from the \HI\ data cube (the
first moment map). The spatial resolution is 58\farcs5 $\times$ 52\farcs6.
The velocity contours
go from  929~\kms\  to 1040~\kms\ in steps
of 5~\kms. A smooth velocity gradient, consistent with that expected from
rotation is seen across the entire system.}
\label{fig:u3672uv3mrov}
\end{figure*}

\begin{figure}
\includegraphics[width=6.0cm,rotate=270]{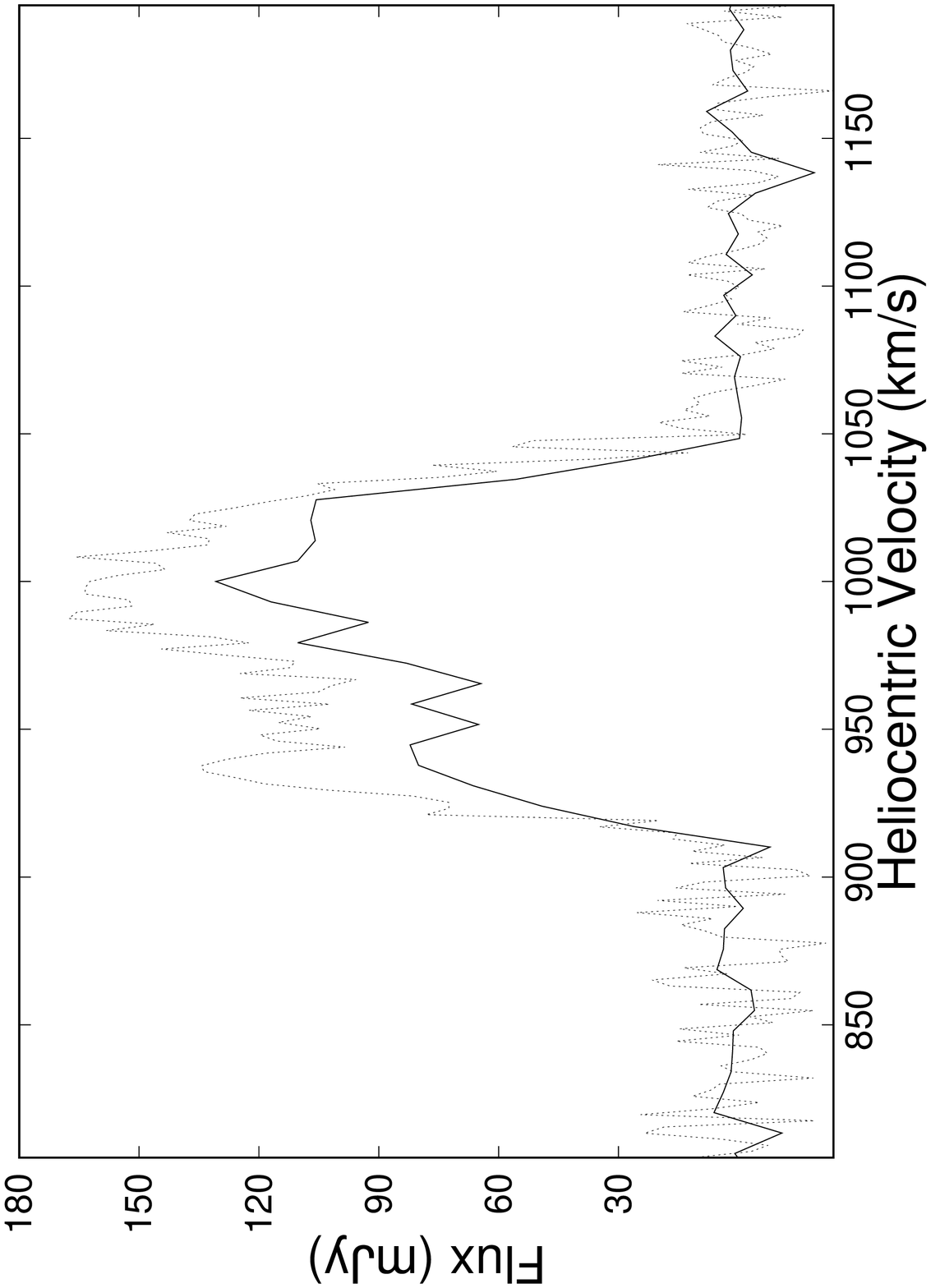}
\caption{The integrated \HI\ spectrum  obtained from the GMRT data cube
  with spatial resolution of 58\farcs5 $\times$ 52\farcs6 and a velocity
  resolution of 6.8~\kms\
(solid line). The dashed line shows the 43-m Green Bank telescope
spectrum \citep{vanzee97,
springob05}. The GMRT spectrum recovers only about 70\% of the flux
detected at 43-m Green Bank telescope. The missing flux, presumably
from a diffuse component
resolved out at the GMRT is fairly uniformly distributed in velocity. }
\label{fig:u3672spc}
\end{figure}

We show in Fig.~\ref{fig:u3672uv3mrov} (left) an overlay of the GMRT \HI\ image
(contours; this image is derived from a data cube with a spatial resolution 
of 58\farcs5 $\times$ 52\farcs6 and a spectral resolution of 6.8~\kms)
on a KPNO 0.9m $B$ band optical image (greyscales; the image is from
\citet{vanzee00}). The \HI\ emission is spread over a region of
$\sim$4.3\arcmin\ (or $\sim 22$~kpc at our assumed distance of 16.9~Mpc). This
is significantly larger than the optical major diameter of $\sim$1.2\arcmin.
The integrated \HI\ spectrum derived from this datacube is shown in
Fig.~\ref{fig:u3672spc}, with the spectrum measured using the Green~Bank
43m telescope \citep{vanzee97, springob05} shown for comparison. The total
flux  measured at the GMRT is $\sim 10.72 \pm 1.1$ Jy~\kms, where
the error bar includes the estimated absolute flux calibration accuracy 
of $\sim 10$\%. The flux measured at the GMRT is only $\sim 70$\% of the
single dish flux. It appears hence that about 30\% of the flux is 
contained in a diffuse component that is resolved out at the GMRT. A 
comparison of the  two spectra in Fig.~\ref{fig:u3672spc} indicates that 
this diffuse emission is fairly uniformly distributed in velocity. It is 
also worth noting that the spectrum appears to have the standard double 
horned shape typical of spiral galaxies. The velocity field observed  at 
the GMRT is shown in Fig.~\ref{fig:u3672uv3mrov}[B]. A smooth gradient can 
be seen from the north-east to the south-west. The entire system hence 
appears to be situated inside an \HI\ envelope with a fairly smooth velocity
gradient, consistent with rotation.

\begin{figure}
\includegraphics[width=6.0cm,rotate=270]{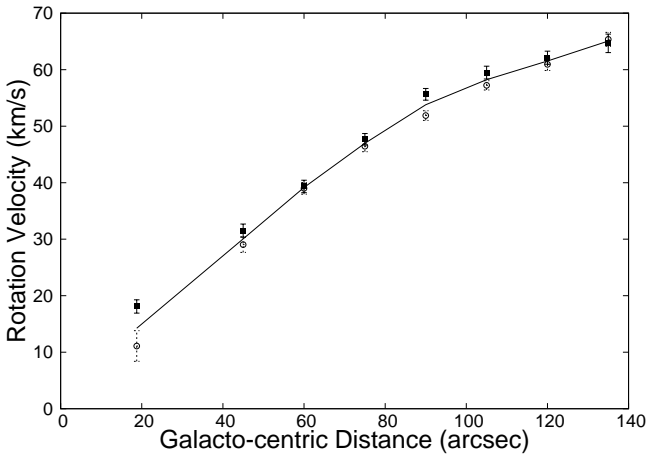}
\caption{The rotation curve for UGC~3672, derived using the GIPSY
program rotcur from the velocity field at a resolution of 
58\farcs5 $\times$ 52\farcs6. The points with the error bars show
the rotation curve as derived for the approaching and receding sides
separately, while the solid line shows the rotation curve derived using
the entire galaxy. There is reasonably good agreement between the
rotation curves derived from the two halves separately.  }
\label{fig:rotcur}
\end{figure}

The rotation curve for the UGC~3672 system was derived from this low
resolution velocity field  using a standard tilted ring  model, as implemented
in the GIPSY\footnote{GIPSY is acronim for Groningen Image Processing SYstem},
task {\tt rotcur}. The initial guess for the centre, systemic velocity and 
position angle were  determined using the {\tt duchamp} package. The initial
inclination was taken to be 60\degr, which is a reasonable estimate from the
\HI\ morphology. These parameters were then determined iteratively using
the rotcur package. The final computed position angle and inclination were 
within  a few degrees of the initial guesses, and the final centre
also coincides with the initial guess to well within the synthesised
beam.  The rotation curve fit by {\tt rotcur } is shown in 
Fig.~\ref{fig:rotcur}. The figure shows the rotation curve fit for the 
approaching and receding half separately (points with error bars) as well 
as the mean rotation curve determined using both the approaching and receding
halves of the galaxy (solid line). The rotation curves derived for the two 
halves separately are in fairly good agreement, reinforcing the visual 
impression that, at this spatial resolution the kinematics are fairly regular.
The velocity at the last measured point of the  rotation curve (which is at 
a galactocentric radius of 135\arcsec, or 11.1~kpc) is 65~\kms.
The corresponding dynamical mass is  $\sim 1.0 \times 10^{10}$~M$_\odot$.

\subsection[]{High Resolution \HI\ images}
\label{ssec:highres}

\begin{figure*}
  \begin{tabular}{p{0.5\textwidth} p{0.5\textwidth}}
    \vspace{0pt} \includegraphics[width=8.0cm]{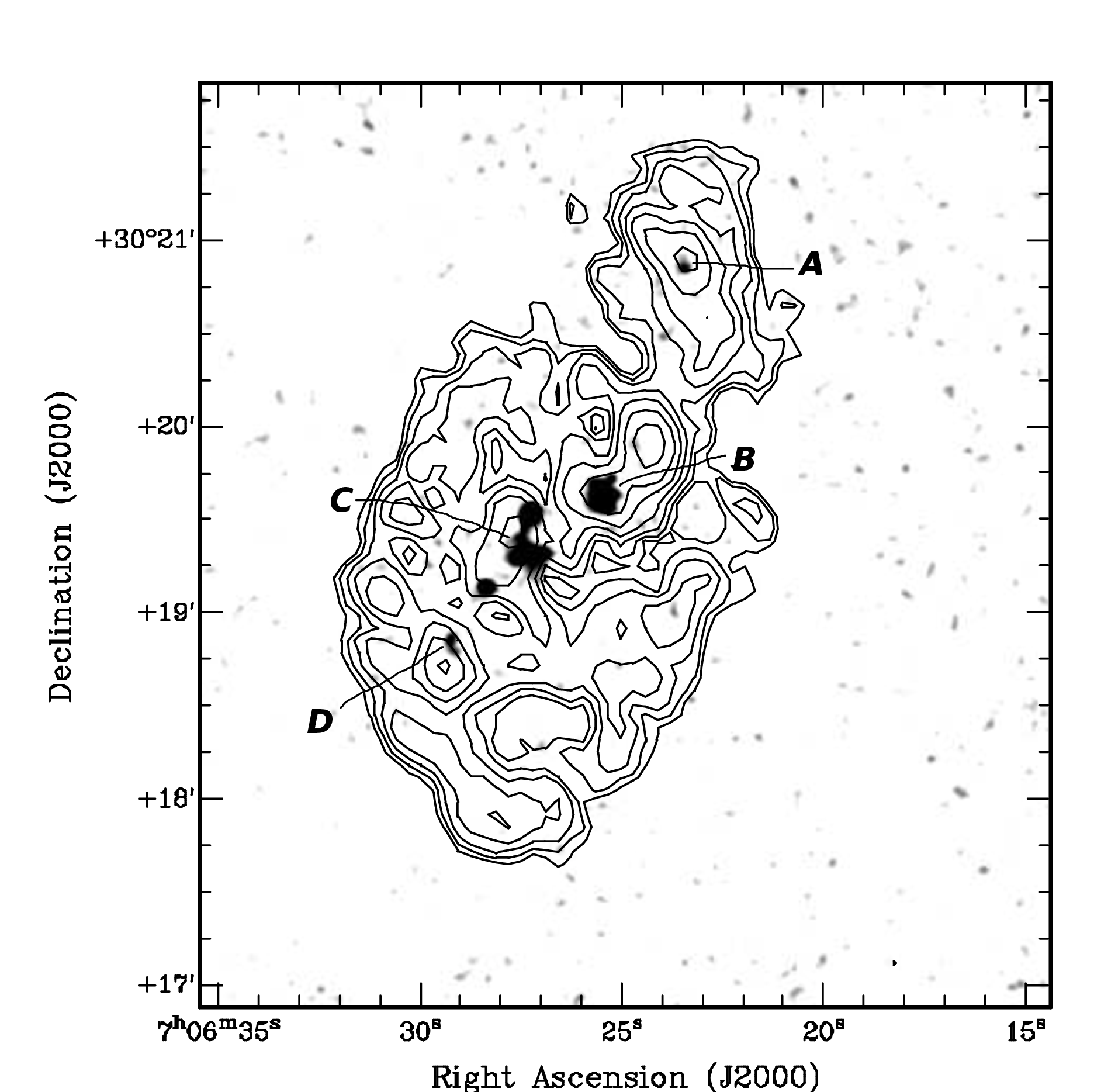} &
    \vspace{-0.7cm} \includegraphics[width=9.5cm]{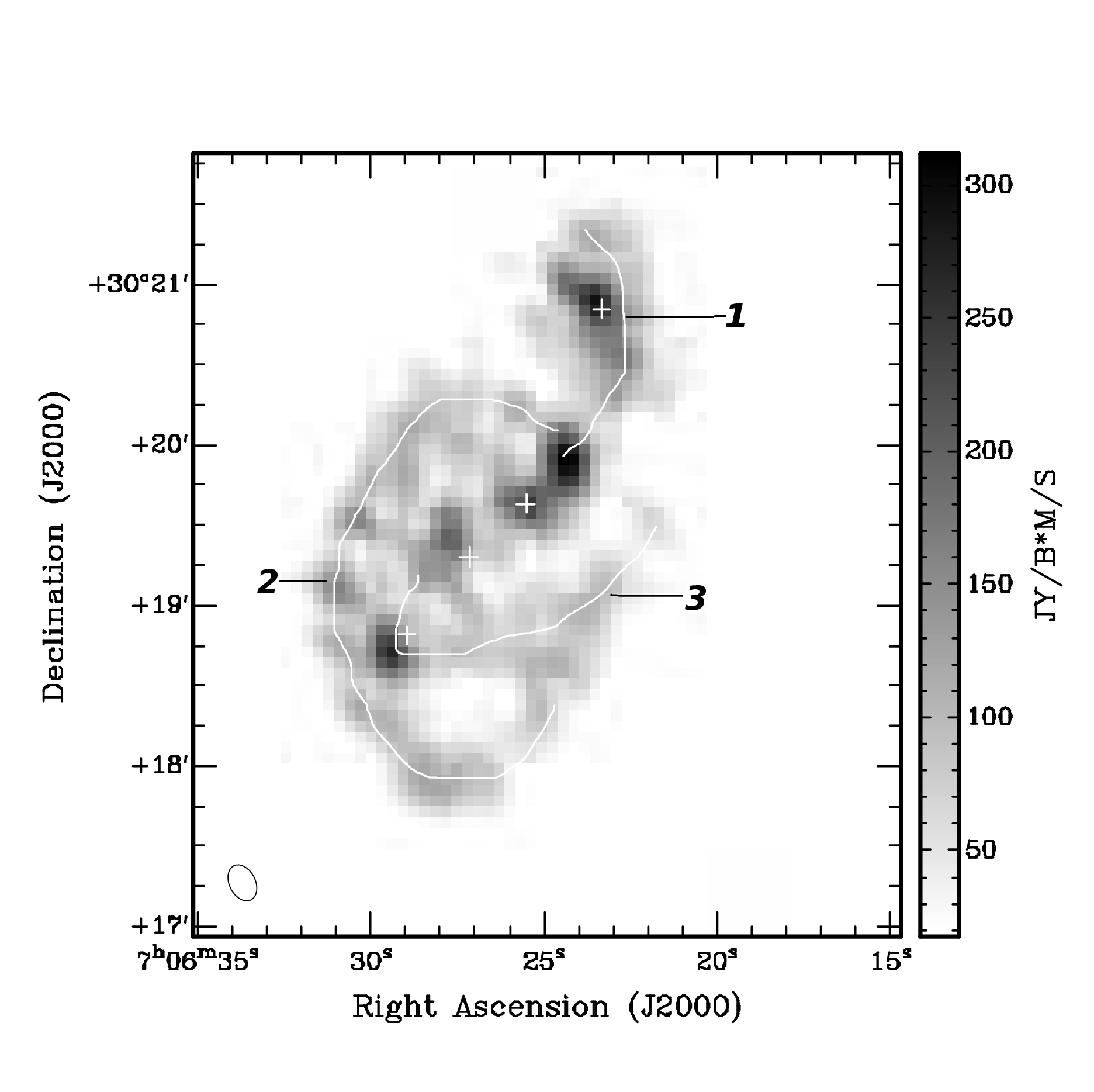}
  \end{tabular}
\caption{{\bf Left:} The \HI\ emission from the UGC~3672 system at a
resolution of 14\farcs6 $\times$ 10\farcs1 (contours) overlaid on the
H$\alpha$ emission (greyscales). The H$\alpha$ data is from \citet{vanzee00},
and has been smoothed to a resolution of 3\arcsec. The \HI\ contours start
at $1.7 \times 10^{20}$~atoms/cm$^{-2}$ and increase in steps of
$\sqrt{2}$. 
{\bf Right:} Greyscale representation of the \HI\ emission at a
resolution of 14\farcs6 $\times$ 10\farcs1. The beam size is shown in
the lower left corner. The locations of the components A to D (going from
north to south) are marked by crosses. Three filamentary structures (1,2,3)
are also marked. As discussed in the text, the velocity varies continuously
along these filaments. 
}
\label{fig:hauv20}
\end{figure*}

Although the low resolution \HI\ emission appears to be somewhat regular,
even from the DSS images of UGC~3672 (not shown) one can see that the 
galaxy has a peculiar appearance, with a diffuse plume which starts from 
the northern end of the brightest emission and  extending westwards. This
feature can also be seen in the KPNO 0.9m $B$-band image
(Fig.~\ref{fig:u3672uv3mrov}). In the $B$-band image one can also see a knot
of emission to the south-east of the main body and faint diffuse emission
about 1\arcmin\ to the north-west. The KPNO 0.9m H$\alpha$ image of the
galaxy  \citep{vanzee00} also shows a faint knot of emission coincident with
the diffuse emission to the north-west seen in the $B$~band image.
\citet{vanzee00} does not comment about this H$\alpha$ knot, but it can be
seen clearly in Fig.~\ref{fig:hauv20} of the current paper. This emission
has also
been identified as arising from a companion dwarf galaxy by \citet{Paper7} 
based on the spectra obtained with the SAO 6m telescope. As discussed further
below, our \HI\ data also supports this
conclusion. We henceforth refer to this galaxy as component A or UGC~3672A.
The peculiar morphology of UGC~3672 along the presence of a nearby companion
galaxy that is embeded in a common \HI\ envelope makes it interesting to
look at the optical and \HI\ images in more detail.

\begin{figure}
\includegraphics[width=8.0cm]{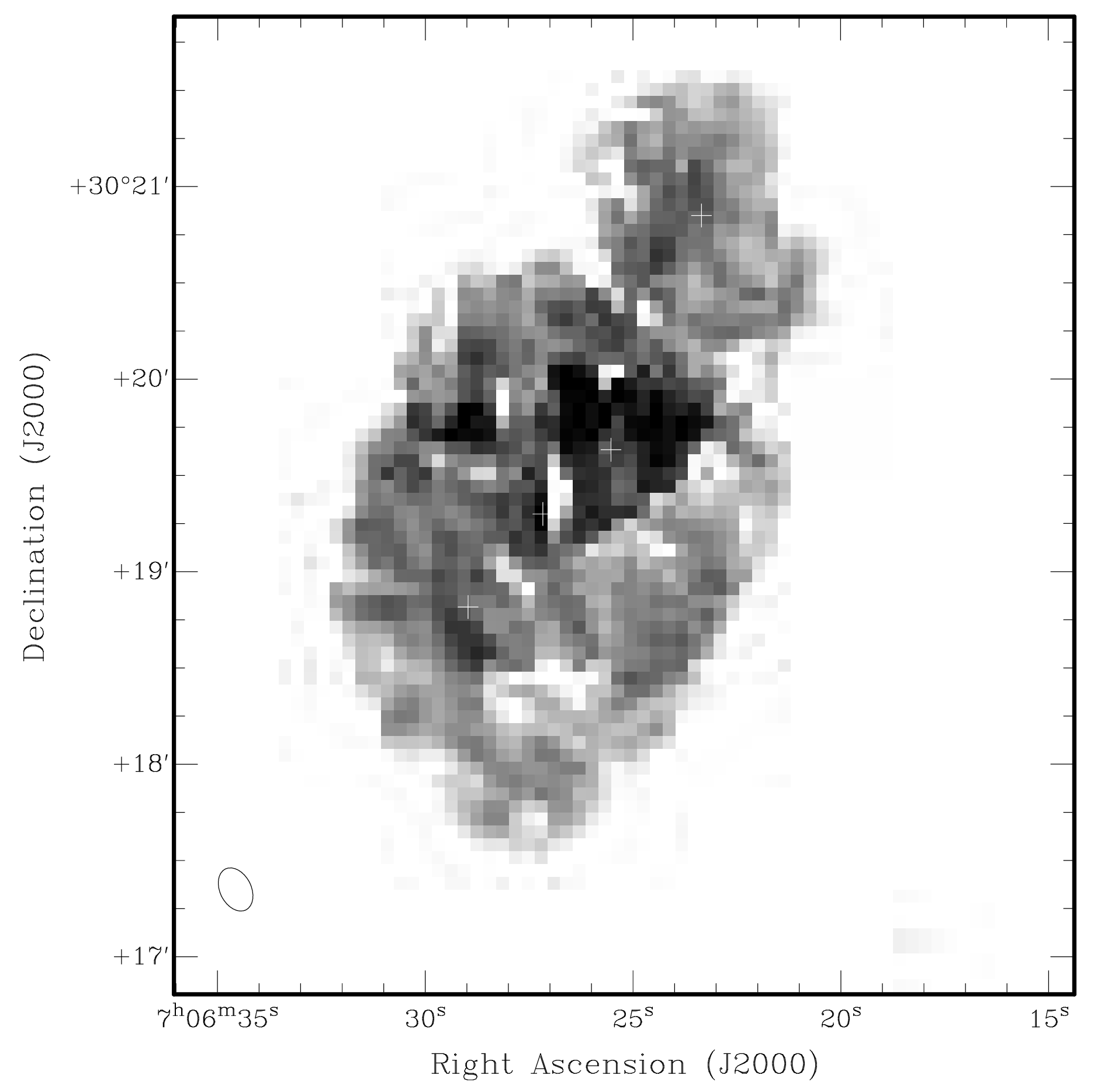}
\caption{ Greyscale representation of the second moment of the \HI\ emission
at a resolution of 14\farcs6 $\times$ 10\farcs1. The beam size is shown
in the lower left corner. The locations of the components A to D (going from
north to south) are marked by crosses. All four components are associated with
local peaks in the velocity dispersion.
}
\label{fig:uv20m2}
\end{figure}

To start with we note that the optical $B$ band image for UGC~3672 itself is 
peculiar, consisting of a diffuse low surface brightness structure which we 
henceforth call component~B, (see the labels in Fig.~\ref{fig:u3672uv3mrov}) 
attached to the north west of a relatively high surface brightness narrow 
emission region (which we henceforth refer to as component~C). The 
H$\alpha$ image (Fig.~\ref{fig:hauv20}) shows knots of bright H$\alpha$ 
emission associated with both of these structures, as  well as a faint knot 
of H$\alpha$  emission south-east of the main galaxy (which we henceforth 
refer to as component~D). In addition, as discussed above, there is also 
H$\alpha$ emission associated with the companion galaxy UGC~3672A.  
The \HI\ emission can also be seen have a filamentary structure (see
Fig.~\ref{fig:hauv20}). As discussed in Sec.~\ref{ssec:nature} these
filaments can be understood as tidal tails arising from the interaction
of extremely gas rich progenitors. We also show in Fig.~\ref{fig:uv20m2}
the second moment (velocity dispersion) of the \HI\ distribution. An increase
in the velocity dispersion near the location of all the four optical
components (A,B,C,D) can be seen.

\subsection{Magnitudes, colours and H$\alpha$ fluxes}
\label{ssec:phot_result}
\begin{table}
\caption{Photometric parameters of UGC~3672 system}
\begin{tabular}{lccc}
\hline
Property               & Comp.~A       & Comp.~B       & Comp.~C+D$^4$ \\
\hline
$B_{\rm tot}$ (SDSS)   & 18.93$\pm$0.15& 17.09$\pm$0.08& 16.32$\pm$0.04    \\
$B_{\rm tot}$ (KPNO)   & 19.09         & 17.15         & 16.26             \\
$M_{\rm B}^1$(mag)     & -12.47        & -14.31        & -15.08        \\
$F_{\rm H\alpha}^2$    & 26$\pm$4.8    & 139$\pm$6.6   & 296$\pm$9.4   \\
$\mu_{\rm B}^3$        & 25.27         & ...           &  ...       \\
$g_{\rm tot}$          &19.39$\pm$0.13 &17.24$\pm0.05$ &16.20$\pm$0.03 \\
$(u-g)_{\rm tot}$      &0.81$\pm$0.33  &\MC {2}{c}{0.86$\pm$0.07} \\
$(g-r)_{\rm tot}$      &--0.08$\pm$0.18&\MC {2}{c}{0.31$\pm$0.04}    \\
$(r-i)_{\rm tot}$      &--0.29$\pm$0.18&\MC {2}{c}{0.16$\pm$0.03}  \\
$(u-g)_{\rm outer}$    & ...           &\MC {2}{c}{0.92$\pm$0.11} \\
$(g-r)_{\rm outer}$    & ...           &\MC {2}{c}{0.23$\pm$0.06}    \\
$(r-i)_{\rm outer}$    & ...           &\MC {2}{c}{0.12$\pm$0.05} \\
\hline
\multicolumn{4}{p{7.6cm}}{
{\bf 1.} Adopted from the SDSS derived $B_{\rm tot}$ and the distance module
of 31.14$^m$, to obtain the robust value of
$M$(\HI)/$L_{\rm B}$. Corrected for galactic extinction $A_{\rm B}=$0.26$^m$.
{\bf 2.} H$\alpha$ in units of 10$^{-16}$ ergs~cm$^{-2}$~s$^{-1}$.
{\bf 3.} mag~arcsec$^{-2}$, average $B$-band surface brightness within the
isophote of SB(B)=26.54$^m$~arcsec$^{-2}$, corrected for $A_{\rm B}=$
0.26$^m$, but not for inclination.  The latter dims the SB by additional
(1.3--1.8)$^m$~arcsec$^{-2}$ for the intrinsic disc axial ratios of
$q$=0.2--0.3. Total magnitudes, $F_{\rm H\alpha}$ and
all colours are not corrected for galactic extinction. {\bf 4.} the light
of components C and D is summed up for $B_{\rm tot}$ and $M_{\rm B}$ and
for parameter $M$(\HI)/$L_{\rm B}$ since for low resolution \HI\ map, used
for the estimate of \HI-fluxes they are situated in a common \HI\ cloud.
}
\end{tabular}
\label{tab:photo}
\end{table}
% E(u-g)=0.069; E(g-r)=0.074; E(r-i)=0.042   A_B=0.264
% A_u=0.31 A_g=0.24 A_r=0.166, A_i=0.124
% A1+A2 u-g_0=0.74+-0.33; g-r_0=-0.15+-0.18; r-i=-0.33+-0.18
% B,C_tot u-g_0=0.79+-0.07; g-r_0=0.24+-0.04; r-i_0=0.12+-0.03
% B,C_per u-g_0=0.85+-0.11; g-r_0=0.16+-0.06; r-i_0=0.08+-0.05
% g_0=19.15; r_0=19.30; g-r_0=-0.15  B_0=19.27
% Ellipse parameters (SDSS) ra=07h06m23.2s  dec=30d20m44.2s, a=24.6",b=8.7"
% SB=26.54

In Table~\ref{tab:photo} we present the  results of the photometry for
components  A, B and C+D. Following \citet{Schlafly11} we take the
$B$ band extinction to be $A_{\rm B} =$ 0.26$^m$. This value was used for
computing both the absolute magnitudes as well as the colour corrections.
In Table~\ref{tab:photo} we show $g_{\rm tot}$ measured from the SDSS images
as well as the derived $B$ magnitudes.  We also show the $(u-g)$, $(g-r)$
and $(r-i)$ colours for the different components. For deriving 
the colours we only use regions with a detectable signal in the $u$ band. 
The $B$ band magnitudes were however estimated over larger apertures using 
the higher signal to noise ratio $g$ and $r$  band images. For components B
and C, the  measured colours are the same within the uncertainties, we hence
tabulate the colours derived for the sum of these components.

We also measured the the $B$-band and H$\alpha$ magnitudes of the different 
components using the KPNO 0.9m images from \citet{vanzee00}. The total $B$ 
magnitude for sum of B, C and D components estimated by us is somewhat fainter 
than that derived for the whole UGC~3672 system by \citet{vanzee00}.
This is likely due to the difference in the sizes of apertures used. Our $B$
magnitude for the sum of B and C+D components is however in accord with 
UGC~3672 total magnitude from \citet{PPK14} and also with that derived 
from our SDSS $g,r$ magnitudes.

A zoom in on component A is shown in Fig~\ref{fig:u3672A}. One can see
that the \HI\ emission peaks near the star forming region, and that in
addition to the star forming knot, there is also a bright foreground
star projected on to the diffuse low surface brightness optical emission.
The foreground star was masked out in order to compute the total magnitude.
As before for deriving  the colours we only use regions with a detectable 
signal in the $u$ band. The $B$ band magnitudes were however estimated over 
larger apertures using  the higher signal to noise ratio $g$ and $r$  band 
images.

\section[]{Discussion}
\label{sec:dis}

\subsection{The nature of UGC~3672}
\label{ssec:nature}

The low resolution HI images suggest that we are dealing with a somewhat 
clumpy rotating gas disk, a not unusual situation for dwarf galaxies 
\citep[see e.g.][]{begum08}. The optical images however show that the
system appears to be disturbed, and as we discussed above, the \HI\
data indicate that UGC~3672 is undergoing an interaction with a much
fainter gas rich companion galaxy which we have named UGC~3672A. 

\begin{figure*}
\includegraphics[width=18.0cm]{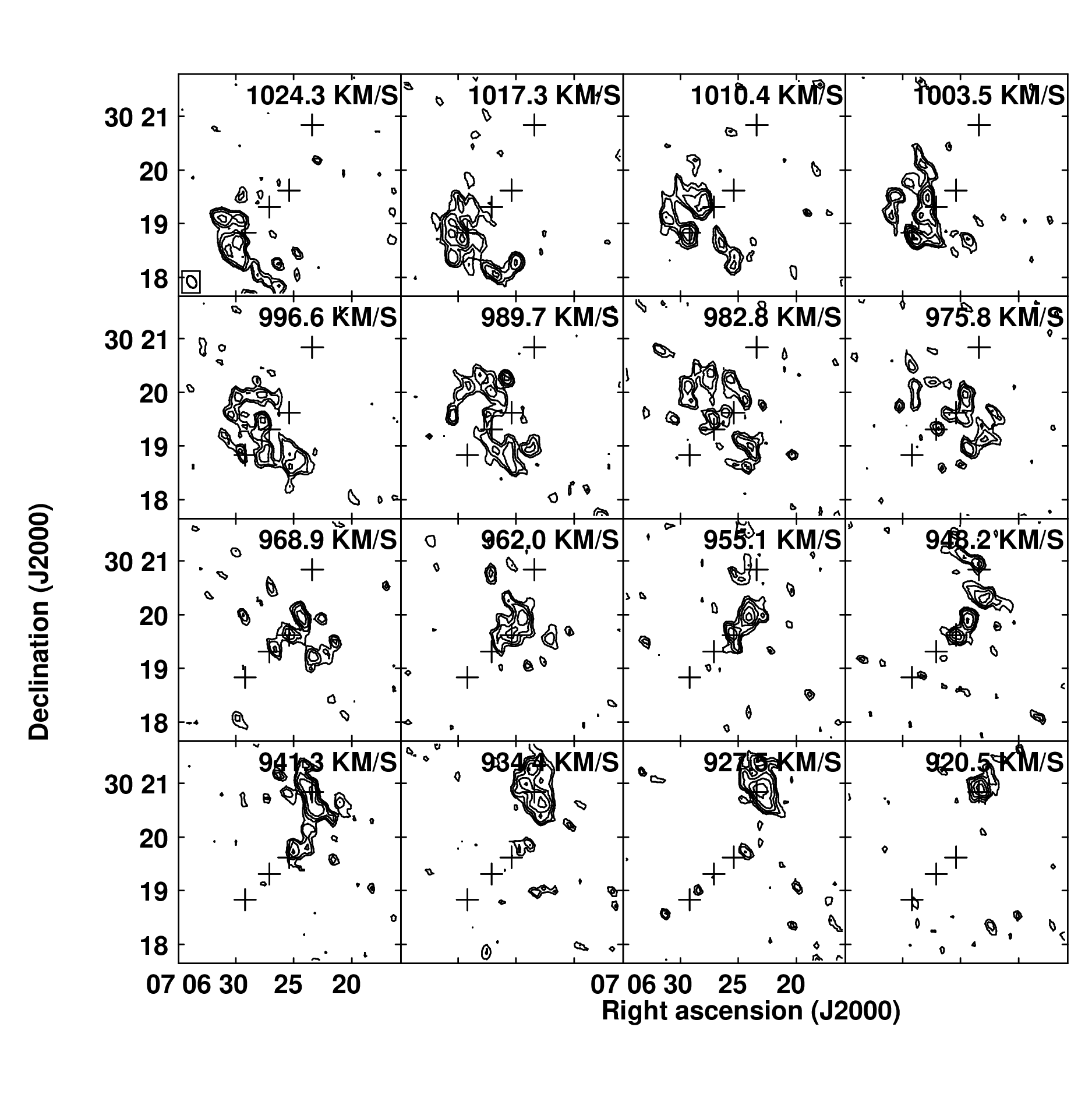}
\caption{ Channel maps of the \HI\ emission from the U3672 at
$14^{``}\times 10^{''}$. These channels do not cover most of the emission,
and this velocity range was selected to clearly bring out the velocity
continuity along the filaments marked in Fig.~\ref{fig:hauv20}. The
contours are logarithmically equally spaced between 2~mJy/Bm and 8~mJy/Bm.
The channel width is $\sim 6.9$~ km/s. The crosses mark the centres of the
components A,B,C,D going from north to south. The beam size is shown in
the lower left corner of the first panel.}
\label{fig:u3672uv20ch}
\end{figure*}

We now look at the UGC~3672 system as a whole. We show in
Fig.~\ref{fig:hauv20} a greyscale representation of the integrated
\HI\ at $\sim 14^{''}$ resolution. The locations of the components A, B,
C and D are also shown using crosses (with A being the northernmost and D the
southernmost). Peaks in the \HI\ emission are seen near all of the components,
although the \HI\ peak near the brightest optical component (C) is
comparatively less bright. The filamentary structure of the HI distribution
can also be seen. To guide the eye and also to correlate this image with the
channel maps in Fig.~\ref{fig:u3672uv20ch}, three filaments labelled 1,2,3
have also been demarcated. The continuity of the velocity fields along these
three filaments can be seen in Fig.~\ref{fig:u3672uv20ch}. The centres of
the four components A,B,C and D are once again marked with crosses. At the
highest velocity channel (1024.3~km/s) the south-west extreme of filament~3
can be seen. In the lower velocity channels one can see that the emission
from this filaments moves towards the north-east, passes through component~D
and then continues to the east of the main optical emission (i.e. the line
joining the central two crosses) and finally terminates (at a velocity
$\sim 975.8$~km/s) slightly to the north of component~B. Emission from
filament~2 starts a little south of component~C (at a velocity
$\sim 1003.5$~km/s) and it too overlaps with region~D, before continuing
to the north-west of the main optical emission and terminating (at
velocities $\sim 962$~km/s). Filament~1 can be seen starting from the
north of component~B (at velocities $\sim 948.2$~km/s) and continues
northward until it terminates near component~A (at velocities
$\sim 920.5$~km/s).

In view of all of this, we propose the following model for this system.
UGC~3672 itself represents the merger of two progenitors, viz. the
components B  and C. Tidal interaction between these, as well
as the third galaxy UGC~3672A  has resulted in the filamentary structures, 
which are essentially tidal tails and arms. As we saw above, the star forming
component D is associated with the intersection of two of these tidal
tails. A possible alternate scenario is one where there are only two
galaxies, viz. UGC~3672 and the companion UGC~3672A. In such a scenario
one would require that the material that forms component B has been tidally
stripped from the main body of UGC~3672 by the companion UGC~3672A. This seems
an unlikely scenario given that UGC~3672A is significantly less massive
than UGC~3672. Support for our conjecture that component B was originally
a distinct system is also provided by the H$\alpha$ kinematics from
Fabri-Perot scanning interferometer observations by \citet{moiseev14}. 
\citet{moiseev14} notes that the kinematics of the system is highly
disturbed, with the kinematical position angle changing by almost 90\degr\
as one moves from the centre outwards (more or less coincident with going
from radii that correspond to component C to radii including components
C and B). Based on the H$\alpha$ kinematics \citet{moiseev14} speculate
that this  is possibly because this object is a merger remnant, in line
with our suggestion that C and B were once distinct entities.

\begin{figure*}
 \begin{tabular}{p{0.5\textwidth} p{0.5\textwidth}}
   \vspace{0pt}\includegraphics[height=9.5cm]{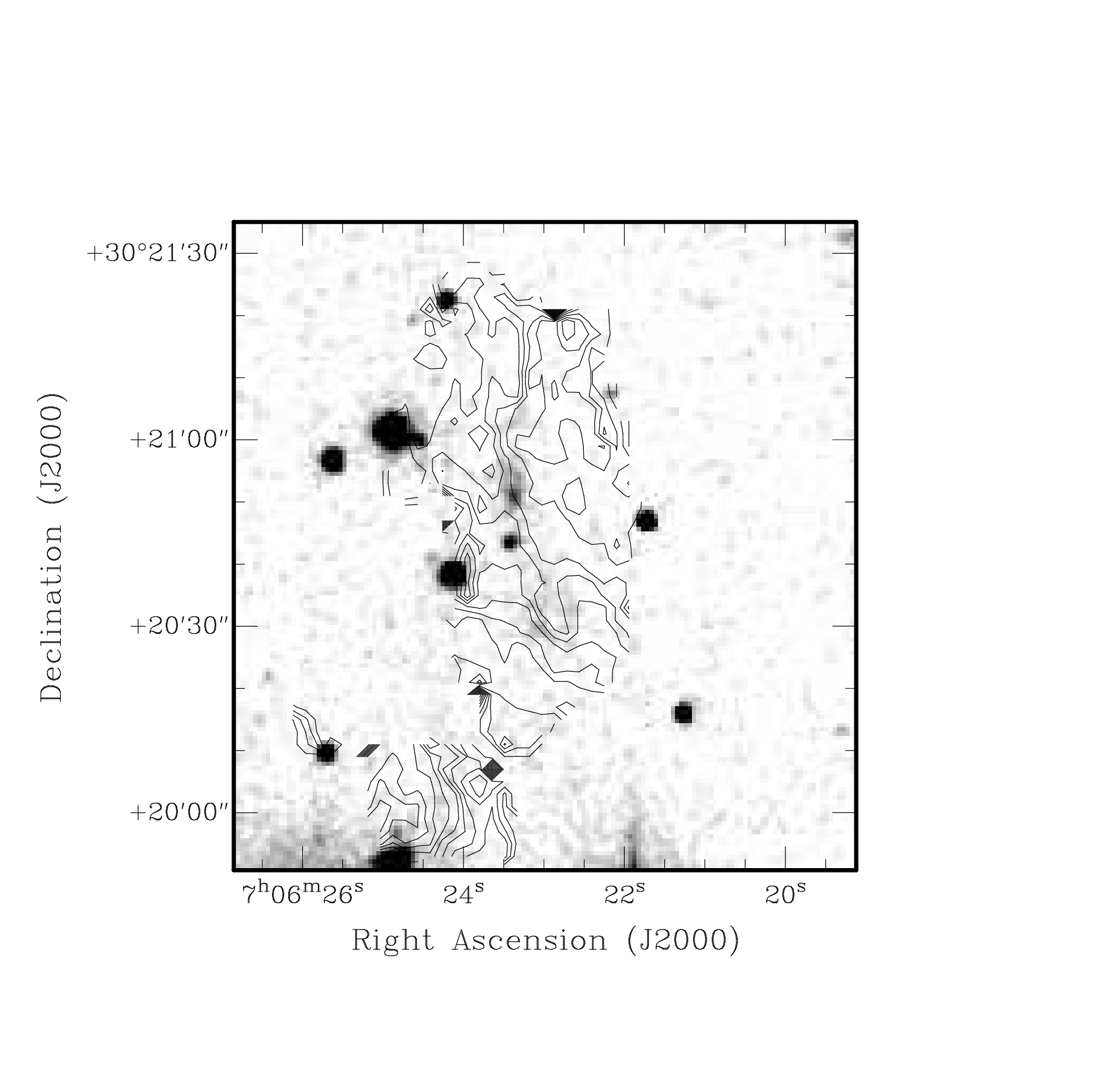} &
   \vspace{0.7cm}\includegraphics[height=8.0cm]{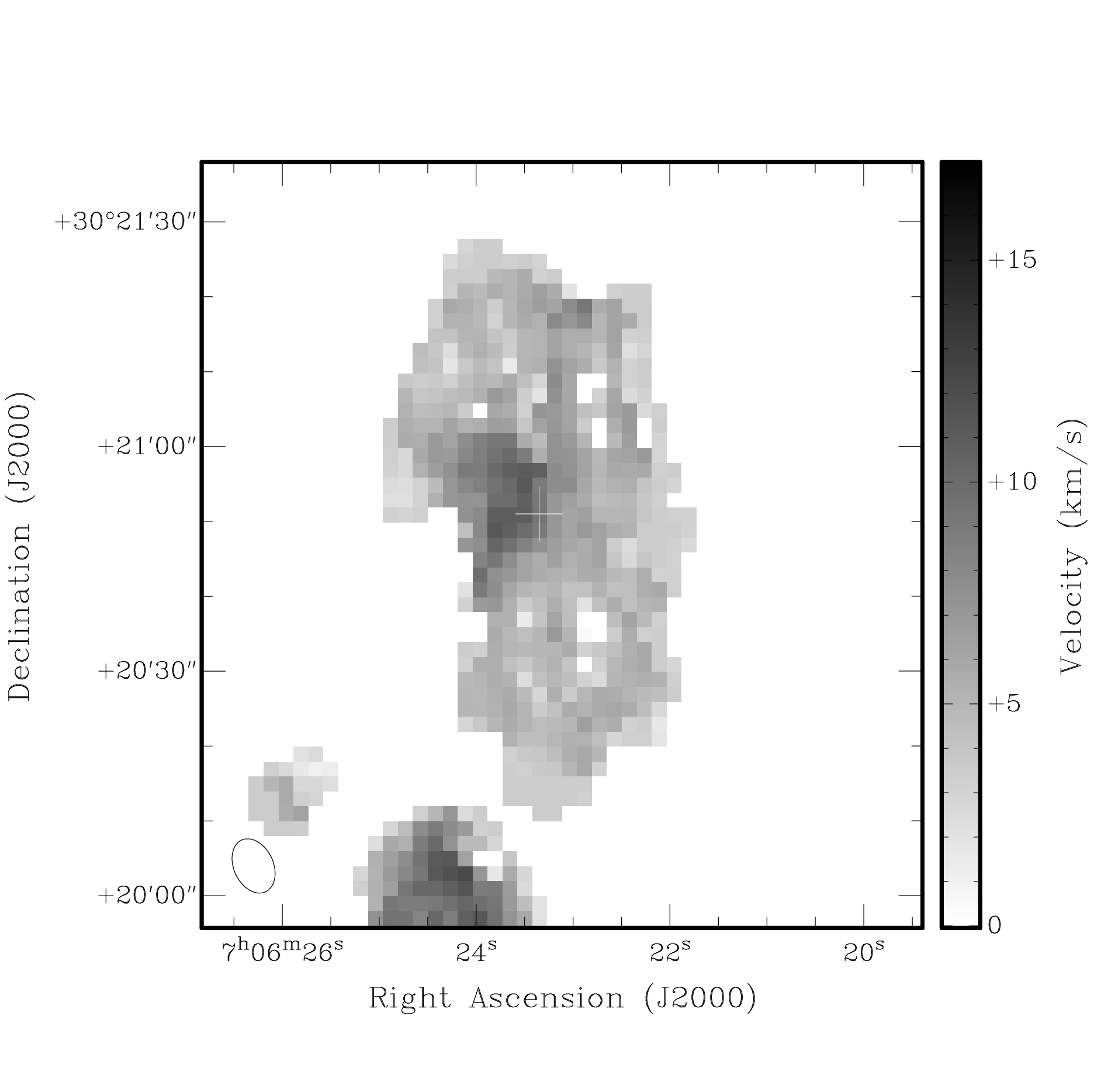}
 \end{tabular}
 \caption{ {\bf left} An overlay of the \HI\ velocity field at 8\farcs1
   $\times$ 5\farcs5 (contours) on the KPNO 0.9m $B$-band image of UGC~3672A.
   The overall North-South extent of the very low surface brightness
   optical emission is $\sim$40\arcsec. A faint foreground star is also
   seen superposed on the diffuse optical emission. A change in the
   direction of the gradient of the \HI\ velocity field around U3672A
   can also be seen.
   {\bf right} An greyscale representation of the second moment of
   the \HI\ distribution at 8\farcs1 $\times$ 5\farcs5. The beam size
   is shown in the lower left corner. An increase in the velocity
   dispersion around U3672A can be seen. 
 }
\label{fig:u3672A}
\end{figure*}

We have so far been treating component A as a separate galaxy. Is there
any kinematic evidence for this? We show the velocity field as well
as the second moment map of the \HI\ emission around U~3672A in
Fig.~\ref{fig:u3672A}. As can be seen the velocity gradient (which
corresponds to that along filament~1 in Fig.~\ref{fig:hauv20}) starts out
in the north-south direction, but near component~A it the gradient
changes direction and is aligned along the east-west direction. The
second moment map Fig.~\ref{fig:u3672A} (right panel) also shows an
increase in the velocity dispersion around component~A. As such,
component~A, appears to be a kinematically distinct component,
consistent with our treatment of it as a separate galaxy.

Assuming then that the UGC~3672 system represents the merger of what
was originally a galaxy triplet, it is interesting to determine
the properties of the progenitor systems. We list in Table~\ref{tab:prog}
the $B$ band magnitudes and luminosities as well as the \HI\ fluxes
and masses of the progenitors. We have measured the
fluxes using the AIPS task BLSUM, summing over the regions around each
optical component. We note that the identification of the \HI\ associated
with the different components is somewhat subjective, given the complex
nature of the \HI\ distribution. The measurements given below are hence
somewhat uncertain. The uncertainty is probably least for component~A,
given that it is reasonably well separated from the rest of the emission.

\begin{table}
\caption{Properties of the progenitors in system UGC~3672}
\begin{tabular}{lccc}
\hline
Property                  & Comp.~A       & Comp.~B        & Comp.~C+D\\
\hline
%$B_{\rm tot}^1$ (mag)     & 18.93         & 17.09          & 16.48    \\
$B_{\rm tot}^1$ (mag)     & 18.93         & 17.09          & 16.32    \\
$M_{\rm B}^2$(mag)        & --12.47       & --14.31        & --15.08   \\
$L_{\rm B}^2$(10$^7~L$\sunn) & 1.51       & 8.24           & 16.7   \\
$V_{\rm hel}^3$(\kms)     &936.7$\pm$1.3  &976.5$\pm$1.3   &1015.9$\pm$3.5 \\
S$_{\rm HI}$(Jy~\kms)     &2.38$\pm$0.5   &3.66$\pm$0.7    &4.7$\pm$0.8 \\
S1$_{\rm HI}^4$(Jy~\kms)  &3.41$\pm$0.7   &5.21$\pm$1.0    &6.73$\pm$1.1 \\
M1$_{\rm HI}^5$(10$^{7}~M$\sunn)&25.8$\pm$5 &35$\pm$7      &45$\pm$8    \\
M1$_{\rm HI}$/$L_{\rm B}$($M$\sunn/$L$\sunn) & 17.1 & 4.2  &   2.7      \\
$L_{\rm H\alpha}$ (10$^{37}$ ergs~s$^{-1}$)       & 9.8          & 52.3           & 124.0      \\
SFR $^6$ ($M$\sunn/yr)     & 0.0006        & 0.0031          & 0.0074      \\
\hline
\multicolumn{4}{p{7.6cm}}{
{\bf 1.} As obtained from the SDSS images. 
{\bf 2.} For the assumed here and below distance of 16.9 Mpc, and including a
 galactic extinction of $A_{\rm B}$ = 0.26$^m$ \citep{Schlafly11}.
{\bf 3.} Taken from the $\sim$60\arcsec\ resolution data cube. The
  uncertainties in this and all of the \HI\ quantities listed here include
  also those related to assigning emission to a given component.
{\bf 4.} This is derived from the above row, but assuming
   that the total flux is as measured at 43-m Green Bank telescope (43mGB),
   and that the flux in each component is scaled proportionately.
{\bf 5.} Derived from the fluxes scaled to the  43mGB value.
%{\bf 6.} H$\alpha$ luminosity in units of 10$^{37}$ ergs~s$^{-1}$,
%  corrected for galactic extinction.
{\bf 6.} Star Formation Rate as derived from H$\alpha$ luminosity $L_{\rm H\alpha}$
  with the formula from \citet{Hunter04}.
%SFR(Ha) (in Mo/yr) = 5.96 * 10^(-42) (in erg/s) (Hunter \& Elmegreen, 2004, AJ, 128, 2170)
%    0.0006  0.0031 0.0074
}
\end{tabular}
\label{tab:prog}
\end{table}

Summarizing the properties of the triplet components in Table~\ref{tab:prog},
it is worth emphasizing the extremely large ratio $M$(\HI)/$L_{\rm B}$ = 17.1
(in solar units) for the faint companion LSBD UGC~3672A. This object is
one of a handful of galaxies with very high, well measured 
$M$(\HI)/$L_{\rm B}$ ratios. We note also that  two other very  gas-rich 
LSBDs have been found in the central regions of the Lynx-Cancer void. The 
others two being J0723+3622 and J0723+3624 with $M$(\HI)/$L_{\rm B} \sim$ 10 
and 26, respectively \citep{chengalur13}. Components `B' and `C+D', are
also gas rich having $M$(\HI)/$L_{\rm B}$ of 4.2 and $\sim$2.6, respectively. 
The total $B$=15.87$^m$ that we measure is $\sim 0.44\pm0.20^m$ 
however fainter than that given by \citet{vanzee00} for UGC~3672. 
As discussed above we believe that our values are the appropriate ones.
If however we adopt the \citet{vanzee00} measurement of 
$B$(UGC3672) = 15.43$^m$, then the parameter $M$(\HI)/$L_{\rm B}$ for the sum 
light of components `B' and `C+D' drops from $\sim$3.2 to $\sim$2.1.

\begin{figure*}
\includegraphics[angle=-90,width=8.0cm]{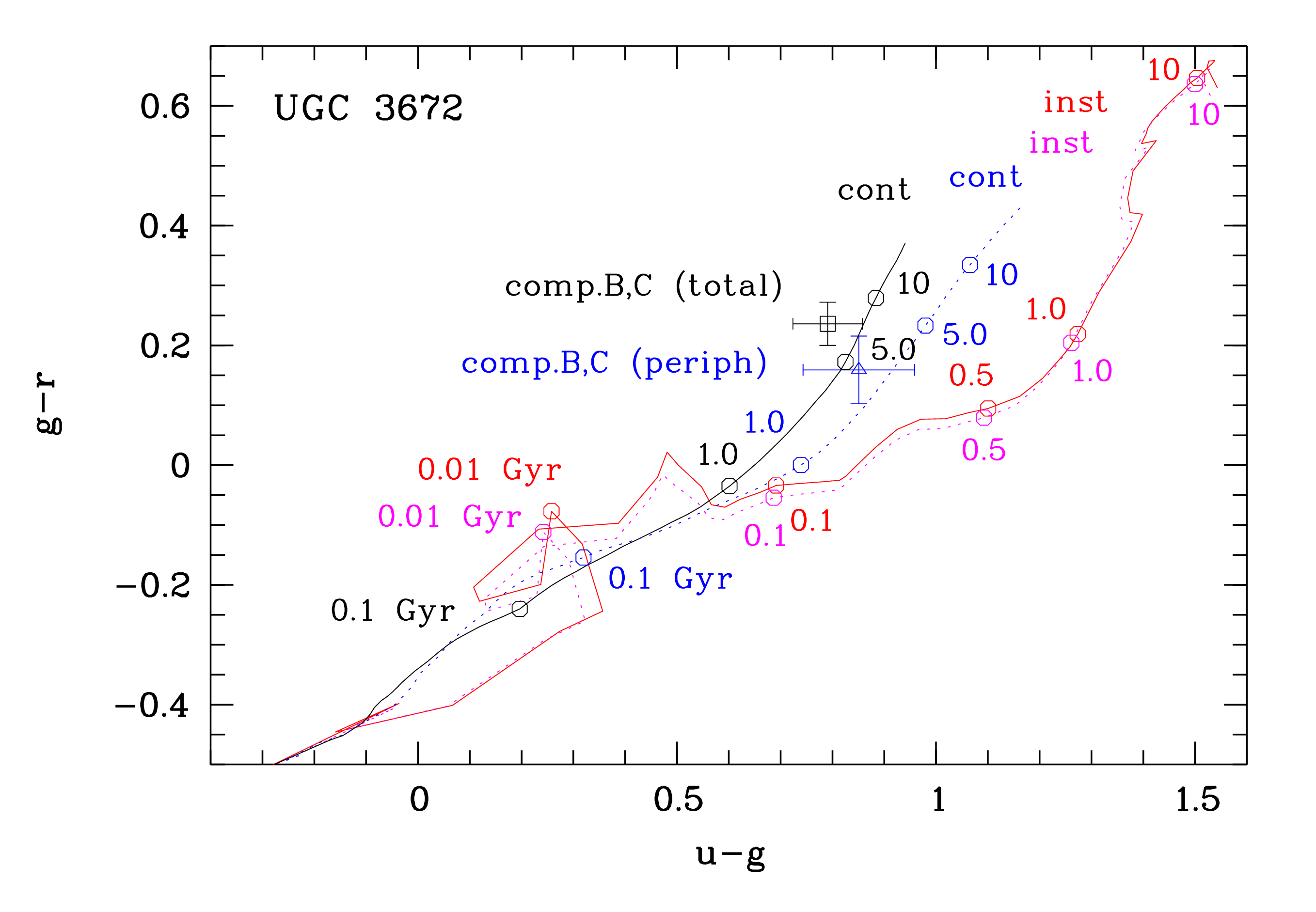}
\includegraphics[angle=-90,width=8.0cm]{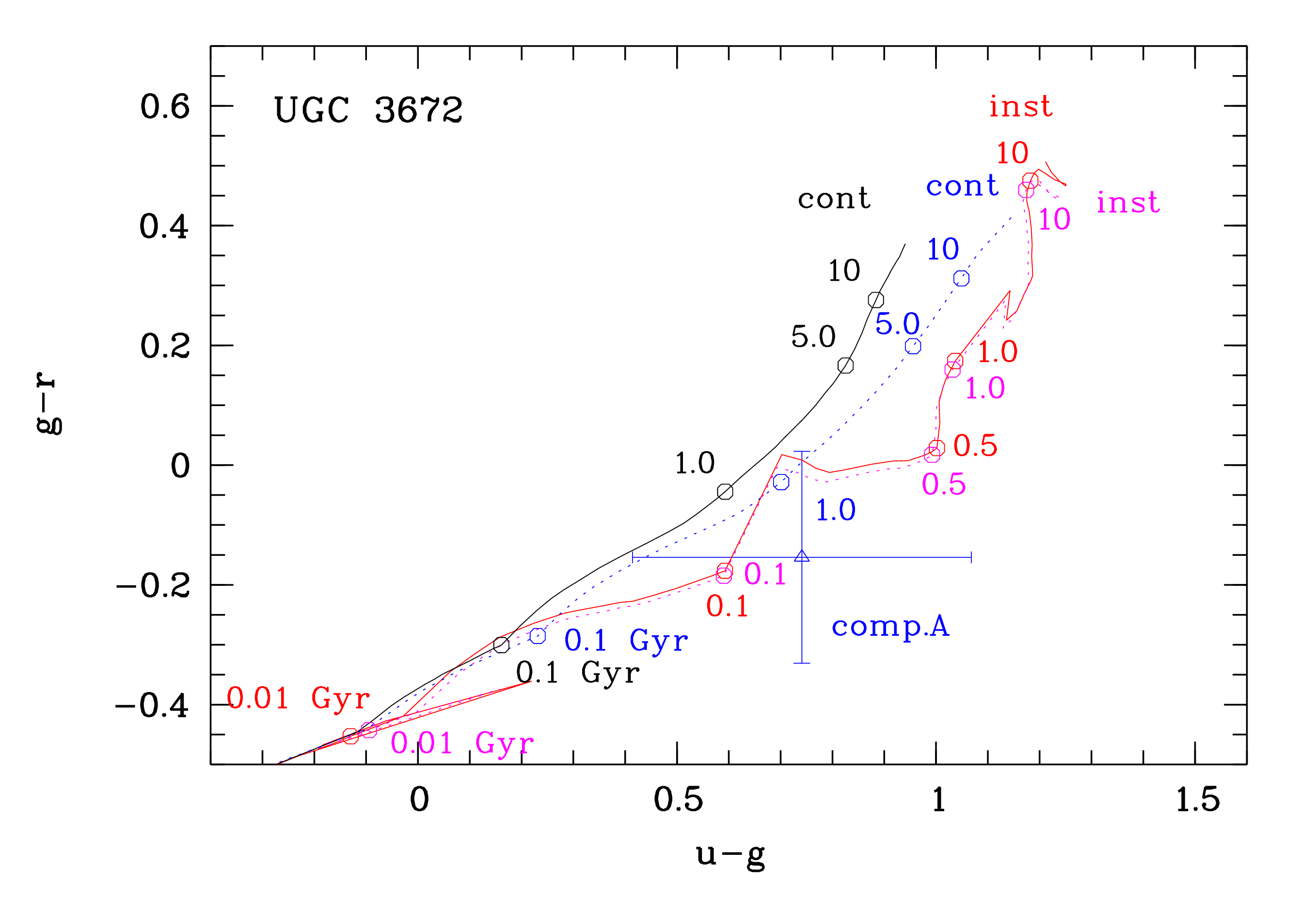}
\caption{Extinction-corrected colours $g-r$ vs $u-g$ with their
$\pm$1$\sigma$ uncertainties as measured on the SDSS DR12
images. {\bf Left panel:} blue triangle - for outer parts of combined
components B and C, black square - colours of B and C integrated light.
{\bf Right panel:} for the total light of component A. They are
overlaid on the PEGASE2 package evolutionary tracks from \citet{FR99}
with tickmarks in Gyr.
Both tracks for continuous SF law with constant SFR ({\it cont}) and the
instantaneous starburst ({\it inst}) are shown as extreme options. See more
details in the text. For components B and C the metallicity of tracks is
$z=0.004$, close to that of HII regions with 12+$\log$(O/H) $\sim$8.0.
For component A we adopt the metallicity of tracks $z=0.0004$, corresponding
to its tentative gas metallicity of 12+$\log$(O/H)=7.0 from Pustilnik,
Perepelitsyna \& Kniazev (2016, MNRAS, submitted).
}
\label{fig:tracks}
\end{figure*}

We also use the observed colours and \mbox{PEGASE2} models with continuous as
well as starburst models to try and constrain the ages of the different
components. In Fig.~\ref{fig:tracks} we show in plot $(g-r)$ vs $(u-g)$ the 
positions of all UGC~3672 system components, with the overlaid \mbox{PEGASE2} 
evolutionary tracks. For the left panel, the adopted value of stellar 
metallicity $z$=0.004 is close to that measured for HII regions in components 
B and C (12+$\log$(O/H)$\sim$8.0). For the very gas-rich and low-mass 
component A, the measured recently value of 12+$\log$(O/H)$\sim$7.0 
\citep{Paper7}, corresponds to a metallicity of
 $z$=0.0004 which is adopted for the evolutionary tracks at the right panel.

Similar to several other faint LSB dwarfs found among the Lynx-Cancer galaxies
by \citet{PPK14}, component A shows very blue $ugr$ colours.
This indicates a small age for its main stellar population, although we note
that the uncertainties are large. Specifically, for component A,
the age range for the nearest model track corresponds to the
instantaneous starburst model with elapsed time from the burst of $\sim$0.1~Gyr.
The track for continuous SF in this region of $ugr$ plot runs quite close
to instantaneous track, and the corresponding age ranges are $T \sim$0.3--1 Gyr,
depending on the assumed form of IMF, Salpeter or Kroupa, respectively.
We note also that the age estimates for component A are much smaller than 
those for the outer parts of components B and C.

%Whether the older
%stellar population exists in this very unusual but rather distant
%void small dwarf,
%we can hope to check only with the next generation giant optical
%telescopes. We seems need to wait for either the ground based ones,
%like E-ELT and TNT, or the new space telescope JWST.

\subsection{UGC~3672 in context}
\label{ssec:context}
 
As mentioned in the introduction, UGC~3672 lies in the central 8\% of
the Lynx-Cancer void volume. For the Lynx-Cancer void as a whole,
the density
of galaxies with $M_{\rm B} \lesssim -14$ is estimated to be $\sim$10\% of
the mean value \citep{chengalur13}. UGC~3672 is also a member of the
Catalog of Nearby Isolated galaxies \citep{karachentsev11}.
UGC~3672 is clearly
located in a highly unusual region. As described above, the properties
of the system are also highly unusual. It appears unlikely that the
fact that the system is highly unusual as well as its environment
is higly unusual is just a coincidence; it appears more reasonable to
conjecture that finding such a system near the center of a void is 
related to the ``cosmic microscope and time machine'' effects associated
with the lowest density regions of the universe. An additional evidence
for this effect is the finding of another unusual system --
very gas-rich dwarf triplet J0723+36, also near the void centre, only at
$\sim$2.2~Mpc from UGC~3672 \citep{chengalur13}.

In this context, it is interesting to note that the UGC~3672 system in some 
ways resembles a scaled down version of well known nearby interacting groups,
e.g., the M81 group \citep{yun94}. In the M81 group also one can see tidal
streamers and bridges connecting the galaxies in the group. The total HI 
mass of the M81 system is however $\sim 5\times 10^9$~$M_\odot$, $\sim$5
times larger than the total HI mass of the UGC~3672 system.  The length 
scales are also similiarly a factor of $\sim$7 larger,  with the total
extent of the M82 system being $\sim 150$~kpc, and the individual tidal 
streamers being $\sim 60$~kpc in length.

It is also interesting to note the ongoing star formation in the
tidal features in UGC~3672, particularly the bright star forming
knot in component 'D' where two separate streamers appear to intersect.
This is similar to what is seen in other larger tidal tails 
\citep[e.g.][]{hibbard05, neff05}.

Star formation in the companion galaxy UGC~3672A is also interesting.
Numerical simulations of star formation triggered by tidal
interactions of gas-rich dwarf systems show that this can lead to bright,
compact central star forming regions, similar to what we see here 
\citep{bekki15a}. Such interactions could also lead to the formation of 
extremely metal defficient (XMD) galaxies \citep{ekta10, bekki08, bekki15b}.
Indeed, the fresh 6-m SAO telescope spectroscopy of the faint HII region
in UGC~3672A reveals an extremely low  gas metallicity (O/H $\sim$ (O/H)\sunn/50;
Pustilnik et al. 2016, submitted). In an alternative scenario, based on the
cold ambient intergalactic gas accretion along the dark matter filaments
\citep[e.g.][and references therein]{sanchez14,sanchez15}, one expects an
asymmetric location of starburst (similar to tadpole galaxies), with the
locally reduced metallicity being due to the mixture of galaxy gas with the
ambient accreted intergalactic gas.

The metallicity of  UGC~3672A is extremely low, and is in fact close to 
the metallicity floor $\sim Z$\sunn/50 found for both the star forming
regions of XMDs and BCGs as well as their outer envelopes 
\citep[][and references therein]{thuan05,izotov12, guseva15}.
The presence of this 'floor' is postulated to reflect pre-enrichment
 by Population III stars of the material out of which these galaxies formed.
This makes it interesting to speculate whether UGC~3672A has undergone most or
all of its star formation fairly recently, i.e. timescales
($T_{\rm SF} \lesssim$~1~Gyr as its $(u-g)$, $(g-r)$ colours 
in Fig.~\ref{fig:tracks} indicate).  Such a very late 'recent' onset of 
SF can be related to the inflow along the void filament to
the pre-existing pair of currently merging components 'B' and 'C'. The tidal
action of the 'massive' DM halo of this pair could induce 
disturbance and related star formation on timescales of $\lesssim$~1~Gyr.
The crucial tests for checks of various alternatives scenarios for
star formation in UGC~3672A would require studies
of the resolved old stellar population with the next generation
giant optical telescopes like E-ELT, TMT and JWST.

Recently \citep{Leoncino_Dw} reported the discovery of another record-low
metallicity dwarf SDSS J094332.35+332657.6 identified with the ALFALFA
\HI-source AGC~198691  Its metallicity (12+$\log$(O/H)=7.02$\pm$0.03 derived 
via the direct method), is close to the mentioned above 'floor'. This faint
($B_{0}$=19.76) extremely gas-rich dwarf with $M$(\HI)/$L_{\rm B}$=6.5 is also 
very blue and probably unevolved. The galaxy is situated at an angular
distance of $\sim$13\arcmin\ from the 3.5$^m$ brighter Lynx-Cancer void 
galaxy UGC~5186 [$D \sim$10.8~Mpc \citep{pustilnik11}]. The heliocentric
velocity of AGC~198691 is $V_{\rm hel}$=514~\kms, only 
35~\kms\ smaller than its brighter counterpart. If we assume that these
two galaxies form a bound system, their angular separation corresponds 
to a linear projected separation of $\sim$40~kpc. Then one could expect 
to see signs of tidal disturbance in the morphology and kinematics 
of the smaller galaxy in deep \HI\ observations.
If AGC~198691 is not bound to UGC~5186, it may reside
$\sim$0.5~Mpc closer and belong to one of the other filaments in
the galaxy distribution identified in Pustilnik et al. (in preparation).  
Thus, this new
record-low metallicity object may have a nature similar to that of UGC3672A, 
emphasising once more the potential of finding atypical low-mass galaxies
in voids.

In contrast to UGC~3672A, star formation is much more wide spread in
the remainder of the system. Metallicity information is available for
two \HII\ regions in the system \citep{vanzee97b}, one each in what we
have designated here as components 'B' and 'C'. Both measurements
agree within the error bars and the average oxygen abundance is
12 + $\log$(O/H) = 8.01$\pm$0.1. This value is very close to
7.99$\pm$0.05, obtained on the independent determination in
\citet{pustilniketal11a}. Both values of O/H are consistent with what is
expected from the luminosity-metallicity relation \citep{ekta10,berg12}.

In the context of its location deep inside a void, the well defined tidal
tails and bridges in the UGC~3672 system, as well as the fact that the 
diffuse gas looks like it is settling into a rotating disk, is interesting.
The role of mergers in determining the morphology of current day galaxies 
has long been an issue of interest. It has been known for decades that 
merger of disk galaxies could result in remnants that resemble ellipticals 
\citep[e.g.][]{toomre72,hernquist92,barnes96}. However, in hierarchical 
galaxy formation models where mergers are expected to be frequent in the 
past, a more crucial question is as to whether disk galaxies can survive 
mergers.

While early simulations indicated that mergers result in the
destruction of disks, it was later shown that for gas-rich disks (the
so called ``wet'' mergers) even major mergers could result in disk-like
morphology \citep{springel05,robertson06,hopkins09a}. Loss of angular 
momentum of the gas which falls to the centre of the galaxy is what 
causes gas poor disks to get completely destroyed. \citet{hopkins09a} argue
that the angular momentum of the gas is lost by tidal torquing between the 
offset gas and stellar disks that are produced in the early stages of the 
interaction. For very gas-rich galaxies this tidal torquing is of less
importance, and most of the gas retains its high angular momentum and 
resettles into a disk at the end of the merger. \citet{bekki08} also presents
the results of numerical simulations of mergers of extremely gas-rich
dwarfs where some gas is driven inwards to fuel a central star
burst, while the rest reforms into a disk. The physical mechanism which
underlies this is however not discussed. 

More recent simulations however show that not all ``wet'' mergers result 
in the re-formation of a disk. \citet{bournaud11} find that the 
effect of the gas  richness on the merger remnant depends critically on 
whether the disk is dominated by clumpy turbulent gas. For disks dominated by
clumpy turbulent gas, mergers do not produce long kinematically coherent tidal
tails nor does the gas reform into a disk. They argue that since $z\sim 1-2$
galaxies appear to be better described as clumpy turbulent disks, it is 
unlikely that mergers of these galaxies could produce a remnant that is 
like the disk galaxies observed today. The morphology and kinematics of 
UGC~3672 indicates that this is not the situation we are dealing with here. 

The final issue that we would like to draw the readers attention to is the
spatial distribution of the galaxies in the system. As argued above, UGC~3672
can be best understood as an interacting triplet of galaxies, with the
galaxies in the triplet having a more or less linear alignment. This
is similar to the case of the triplet found by \citet{beygu13} in their
Void Galaxy Survey, and which they suggest reflects the underlying filamentary
nature of the distribution of galaxies in voids. The kinematical continuity
that we find across the UGC~3672 system is also similar to that noted
in the study of the NGC~672 and NGC~784 groups by \citet{zitrin08}. These
authors suggest that the kinematical alignment arises because the galaxy 
in the groups like along a filament of the large scale structure.

The galaxy triplet discovered earlier near the centre of the Lynx-Cancer void
\citep{chengalur13} also has a more or less linear geometry, although in
this triplet there is no bridge of emission connecting the third
galaxy with the main pair, and although there is velocity continuity 
across the main pair, it does not extend to the third galaxy. The linear
arrangement and velocity continuity of the galaxies in the UGC~3672 system,
along with the fact that the diffuse gas appears to be settling into a
rotating disk, indicate that wet mergers with flow along filaments
is a possible way to produce disk-like systems.

\section{Summary and Conclusions}

We present multi-resolution \HI\ images and optical photometry data
of the galaxy UGC~3672, which is located near the centre of the nearby
Lynx-Cancer void. From the data analysis and their discussion in the
broader context, we draw the following conclusions.

\begin{enumerate}
\item
At low spatial resolution, the \HI\ distribution appears to be in a clumpy
disk with fairly regular dynamics. At this resolution, the \HI\ distribution
and kinematics is consistent with what is typical of dwarf irregular galaxies.
\item
At high spatial resolutions however, the \HI\ distribution is highly unusual,
consisting of a secondary peak, which is identified with a separate,
much fainter dwarf irregular galaxy, as well as several filamentary
structures.
\item
 This faint LSB dwarf UGC~3672A appears extremely gas-rich, with the ratio
$M$(\HI)/$L_{\rm B} =$ 17.1. Besides, its integrated colours $u-g$ and $g-r$,
when compared with the PEGASE2 package evolutionary tracks indicate that
the bulk of the stars in this system have formed recently. This, along with
its extremely low metalicity (O/H of $\sim$(O/H)\sunn/50) means that
UGC~3672A has all the properties expected of unevolved galaxies.
Along with two other extremely gas-rich and blue LSBDs, from another triplet
J0723+36 near the void centre, residing at only $\sim$2~Mpc from UGC~3672
system, this finding provides an additional evidence for the unevolved nature
of small dwarfs near the void centres.
\item
Arguments are presented that UGC~3672 system is best understood as
representing the merger of what was originally a triplet of very gas-rich
dwarf galaxies. We suggest that the location of this highly unusual system 
is not a coincidence but is related to the ``cosmic microscope and time
machine'' effects associated with voids. 
\item
Our observations indicate that wet mergers of galaxies flowing along a
filament is a possible pathway for the production of disk like systems.
\end{enumerate}

\section*{Acknowledgements}
We thank the staff of the GMRT who made these observations possible. The GMRT
is run by the National Centre for Radio Astrophysics of the Tata Institute of
Fundamental Research. The work of SAP and ESE was supported by the
RSCF grant No.~14-12-00965. 
%RFBR grants 14-02-00520 and 15-52-45004.
%ESE is grateful to the RSCF grant
%No. 14-22-00041 for support of the photometric data analysis.
We acknowledge the use of the SDSS DR12 database.
Funding for the Sloan Digital Sky Survey (SDSS) has been provided by the
Alfred P. Sloan Foundation, the Participating Institutions, the National
Aeronautics and Space Administration, the National Science Foundation,
the U.S. Department of Energy, the Japanese Monbukagakusho, and the Max
Planck Society. The SDSS Web site is http://www.sdss.org/.
The SDSS is managed by the Astrophysical Research Consortium (ARC) for the
Participating Institutions. This research has made use of the NASA/IPAC 
Extragalactic Database (NED) which is operated by the Jet Propulsion 
Laboratory, California Institute of Technology, under contract with the 
National Aeronautics and Space Administration.  We are grateful to the
anonymous referee whose comments helped improve this paper.

%===========================================================================

\bibliographystyle{mn2e}
%\bibliography{voidbib}
\bibliography{u3672r2}

\end{document}